\documentstyle{mn2e}
\input psfig.sty 
 \def\be{\begin{equation}}
 \def\ee{\end{equation}}
\def\solmas{{M$_\odot$}}
\def\simless{\mathbin{\lower 3pt\hbox
   {$\rlap{\raise 5pt\hbox{$\char'074$}}\mathchar"7218$}}}   
\def\simgreat{\mathbin{\lower 3pt\hbox
   {$\rlap{\raise 5pt\hbox{$\char'076$}}\mathchar"7218$}}}   
\def\etal{{\rm et al.}}

\def\solmas{{M$_\odot$}}
\def\solm{{M_\odot}}

\def\apj{{ApJ}}

\def\mnras{{MNRAS}}

  \makeatletter
  \ifx\CUP@mtlplain@loaded\undefined  
  \makeatother  
    
  \else
    
  \fi

\loadboldmathitalic 
\loadboldgreek      
  
\makeatletter
\ifx\CUP@mtlplain@loaded\undefined  
  \newfont\bit{cmbxti10 at 9pt}
\else
  \newfont\bit{mtbxti10 at 9pt}
\fi
\makeatother

\def\LaTeX{L\kern-.36em\raise.3ex\hbox{a}\kern-.15em
    T\kern-.1667em\lower.7ex\hbox{E}\kern-.125emX}

\newcommand{\gsim}{\mathrel{\hbox{\rlap{\lower.55ex \hbox {$\sim$}}
                   \kern-.3em \raise.4ex \hbox{$>$}}}}
\newcommand{\lsim}{\mathrel{\hbox{\rlap{\lower.55ex \hbox {$\sim$}}
                   \kern-.3em \raise.4ex \hbox{$<$}}}}

\title[Formation of brown dwarfs] {Gravitational fragmentation and the formation of brown dwarfs in stellar clusters}
\author[I. A. Bonnell \etal]
  {Ian A. Bonnell$^1$\thanks{E-mail: iab1@st-and.ac.uk}, Paul Clark$^2$ and Matthew R. Bate$^3$ \\
$^1$ SUPA, School of Physics and
  Astronomy, University of St Andrews, North Haugh, St Andrews, Fife,
  KY16 9SS. \\
$^2$  Institut fuer Theoretische Astrophysik, Albert-Ueberle-Str. 2, 
69120, Heidelberg, Germany \\
$^3$ School of Physics, University of Exeter, Stocker Road, Exeter, EX4 4QL \\ }

\date{\today}

\begin{document}

\maketitle

\begin{abstract}
We investigate the formation of brown dwarfs and very low-mass stars through the
gravitational fragmentation of infalling gas into stellar clusters. The gravitational potential of a forming stellar cluster provides the focus that attracts gas from the surrounding molecular cloud. Structures
present in the gas grow, forming filaments flowing into the cluster centre. These filaments attain high gas densities due to the combination of the cluster potential and local self-gravity. The resultant Jeans
masses are low, allowing the formation of very low-mass fragments. The tidal shear and high
velocity dispersion present in the cluster preclude any subsequent accretion thus resulting in the formation of  brown dwarfs or very low-mass stars.  Ejections are not
required as the brown dwarfs enter the cluster with high relative velocities, suggesting that
their disc and binary properties should be similar to that of low-mass stars. This mechanism requires the presence of a strong gravitational potential due to the stellar cluster implying that brown dwarf formation should be
more frequent in stellar clusters than in distributed populations of young stars. Brown dwarfs formed in 
isolation would require another formation mechanism such as due to turbulent fragmentation.
\end{abstract}


\section{Introduction}

Brown dwarfs,  having masses less than 0.08 \solmas, are seen to be nearly as frequent as
stars and to have many of the same properties during their youth such as circumstellar discs, binary companions, and chromospheric activity (Burgasser \etal\ 2007; Luhman \etal\ 2007; Scholz, Jawyawardhana \& Wood 2006).
There have been several proposed mechanisms to explain the origin of brown dwarfs (Whitworth \etal\ 2007). 
These have involved such diverse physical processes as turbulent fragmentation, disc fragmentation,
the ejection of stellar embryos and the photo-evaporation of a collapsing prestellar core. These
different mechanisms all rely in  forming a hence high gas density
in the prestellar cores and thus a  low Jeans mass, and/or in the halting of accretion
once a low-mass fragment has formed. 

Turbulent fragmentation (Padoan \& Nordlund 2002) envisions that strong, magnetic shocks produce high density
post-shock gas which will have low Jeans masses and therefore form brown
dwarfs directly from the turbulence (Padoan \& Nordlund 2004). One of the potential difficulties with this model is that
a straight shock only compresses the gas in 1-D. This does not
affect the overall gravitational radius of the gas  and thus has a negligible effect
of the Jeans mass as the gravitational and thermal energies are basically
unchanged (Elmegreen \& Elmegreen 1978; Lubow \& Pringle 1993;  Clarke 1999). Several roughly coincident shocks are required in order to get 3-D compression
and a reduction in the Jeans mass. The turbulent compression model also neglects any 
residual internal turbulence generated from the shocks (Clark \& Bonnell 2005).

Disc fragmentation (Bate, Bonnell \& Bromm 2002; Whitworth \& Stamatellos 2006; Goodwin \& Whitworth 2007; Stamatellos \etal\  2007) occurs when a massive circumstellar disc is unstable to gravitational
fragmentation, potentially induced by a stellar fly-by (but see Clarke \etal\ 2007). 
The disc provides the high density material such that
the Jeans mass is necessarily low. Post formation accretion from the massive disc onto the brown
dwarf could increase the mass significantly while forming brown dwarfs in single systems
is more difficult. 

Ejection of newly formed fragments in multiple systems (Reipurth \& Clarke 2001; Bate, Bonnell \& Bromm 2002; Bate \& Bonnell 2005)
can halt any post formation accretion such that the fragment maintains a low mass. This still requires that the
Jeans mass at the point of fragmentation is of order a brown dwarf mass such as occurs in circumstellar discs
and infalling filaments. One of the potential difficulties in this mechanism is that the ejection will tend
to truncate discs at radii of order 10 AU, and likewise disrupt any binaries with separations comparable
or larger to this.

One final mechanism is the photoevaporation of collapsing cores (Whitworth \& Zinnecker 2004). This mechanism 
envisions that a collapsing core in the proximity of an O star will be photo-eroded before it can fully collapse.
The outer layer of the core will be ionised and unbound while adding a pressure term onto the
inner collapsing core ensuring that a lower-mass object is formed. The primary difficulty with this
mechanism is that it would need to be finely tuned such that the radiation field is strong enough to have an 
effect but not so strong such that it would completely destroy the core. Furthermore, it requires
the presence of an ionising source of radiation such that it could only explain brown dwarfs in
the presence of O-stars.

In this paper, we present some recent results on the formation of low-mass stars and brown dwarfs
due to the presence of stellar clusters. We show that  brown dwarfs form in an analogous
manner to low-mass stars, due to the gravitational fragmentation of high density gas
as it infalls into a stellar cluster. The primary differences with our previous work (Bate \etal\ 2002) is 
that we stress how the low fragmentation is reached due to gravitational compression, and that
subsequent accretion is limited due to the high virial velocities in clusters such that
ejections are not needed.  Our calculations are presented in \S 2
while the primary results are presented in \S 3. Section 4 dissects the role of the stellar
cluster in forming brown dwarfs while \S 5 presents some observational signatures
of the process described here.

\section{Calculations}

\begin{figure*}
\centerline{\psfig{figure=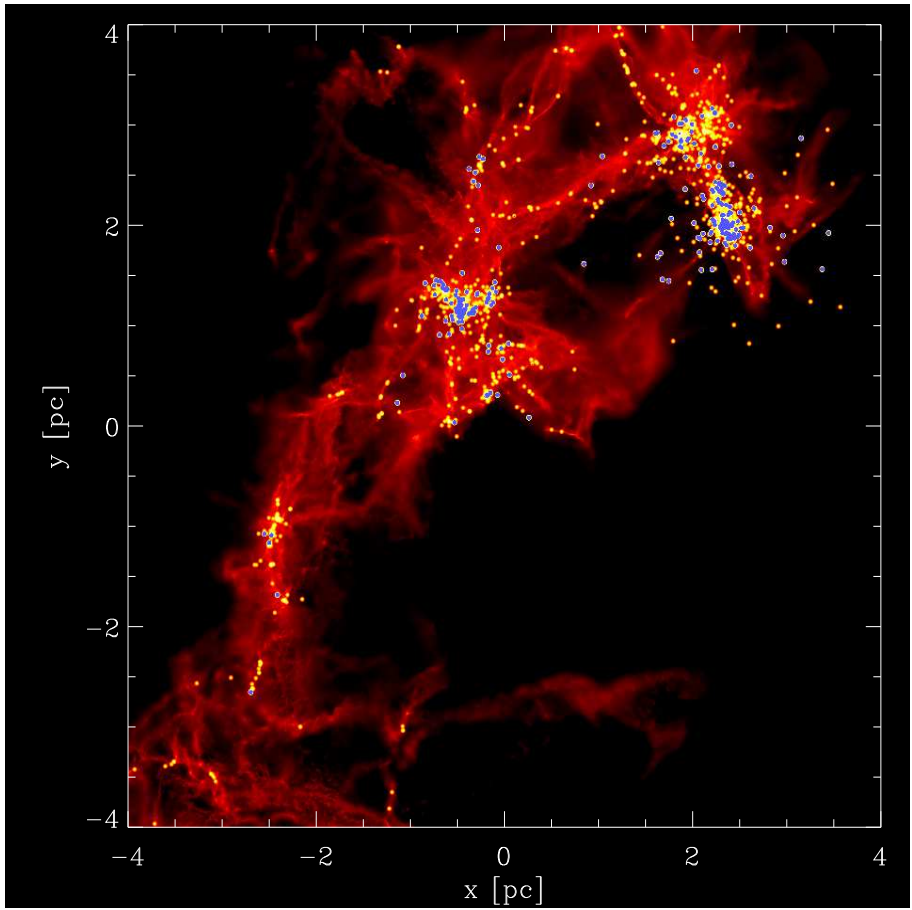,width=12.truecm,height=12.truecm}}
\caption{\label{BDsimage}  The final distribution of the gas and stars in the simulation
is shown in the 8 by 8 pc image. The stars are indicated by the yellow filled circles  while the brown dwarfs are indicated
by the filled blue circles with white edges. The brown dwarfs are located primarily in clustered
regions as these provide the necessary physical properties to form low mass objects. The gas column densities are also plotted from 0.01 (black) to 100 (white) g cm$^{-2}$.}
\end{figure*}

The results presented here are based on a large-scale Smoothed Particle Hydrodynamics (SPH)
simulation of a cylindrical $10^4$ \solmas\ molecular cloud 10 pc in length and 3 pc in cylindrical diameter. We have chosen an elongated cloud rather than the more standard spherical
cloud as most molecular clouds are non-sperhical and commonly elongated (e.g. Orion A).
This also allows for the physical properties to be varied along the cloud in a straightforward manner.
The cloud has a linear density gradient along its major axis with maximum/minimum values,
at each end of the cylinder,
$33$ percent high/lower than the average gas density.
The gas has internal turbulence following a Larson-type $P(k) \sim k^{-4}$
power law and is normalised such that the total kinetic energy balances the total gravitational energy in the cloud.
The density gradient then results in one end of the cloud being over bound (still super virial) while the other
end of the cloud is unbound. 

The cloud is populated with 15.5 million SPH particles on two levels,
providing high resolution in regions of interest. We initially performed a lower resolution
run with  5 million SPH particles producing an average mass resolution of $0.15 \solm$. Upon
completion of this low resolution simulation, we used three criteria to identify the regions that required higher resolution. This included the  particles which formed sinks, and those that were accreted onto sinks. It also included particles which attained sufficiently high density such that their local Jeans
mass was no longer resolved in the low-resolution run. All of these particles were identified and 
from the initial conditions of the low resolution run, they were spit into 9 particles each to create the initial conditions for the high resolution simulations. This 
particle splitting was performed on the initial conditions to ensure that the physical quantities of mass, momentum,  
energy  and the energy spectrum were preserved. Note that the particle splitting does not introduce finer structure in the turbulent energy spectrum. This produced a mass resolution for the regions involved in star formation
of $0.0167 \solm$, sufficient to resolve the formation of brown dwarfs, equivalent to a total
number of  $4.5 \times 10^7$ SPH particles. The equation of state (below) was specified in order to ensure that the Jeans mass in the higher resolution run did not descend below this mass resolution.

This simulation was rerun from the beginning
to ensure that the particle splitting did not affect the ongoing evolution. 
Particle splitting results in a marked increase in resolution without unmanageable computational costs (Kitsionas \& Whitworth 2002,~2007). Note however some of the unsplit particles, which  in the low resolution run neither exceeded their Jeans mass limit 
nor became involved in the star formation, did get accreted by the additional stars in the high resolution run. This is to be expected as their are now additional locations of star formation not present in
the low resolution run and these additional sinks will necessarily accrete unsplit particles.

The simulation follows a barotropic equation of state of the form:
\begin{equation}
P = k \rho^{\gamma}
\end{equation}
 where
\begin{equation}
\begin{array}{rlrl}
\gamma  &=  0.75  ; & \hfill &\rho \le \rho_1 \\
\gamma  &=  1.0  ; & \rho_1 \le & \rho  \le \rho_2 \\
\gamma  &=  1.4  ; & \hfill \rho_2 \le &\rho \le \rho_3 \\
\gamma  &=  1.0  ; & \hfill &\rho \ge \rho_3, \\
\end{array}
\end{equation}
and $\rho_1= 5.5 \times 10^{-19} {\rm g\ cm}^{-3} , \rho_2=5.5 \times 10^{-15} {\rm g\ cm}^{-3} , \rho_3=2 \times 10^{-13} {\rm g\ cm}^{-3}$.
 
The initial cooling part of the equation of state mimics the effects of line cooling and ensures that the
Jeans mass at the point of fragmentation is appropriate for characteristic stellar mass (Larson 2005; Jappsen \etal\  2005; Bonnell, Clarke \& Bate 2006).
The $\gamma=1.0$ approximates the effect of dust cooling (Larson 2005) while the $\gamma=1.4$ mimics the effects of when the collapsing
core is optically thick to IR radiation, although its location at $\rho= 5.5 \times 10^{-15} {\rm g cm}^{-3}$, at lower densities than is typical, is in order to ensure that
the Jeans mass is always fully resolved and that a single self-gravitating fragment is turned into a sink particle. A higher critical density for this optically-thick phase where heating occurs would likely result in an increase in the numbers of brown dwarfs formed. the physical processes described would be unchanged. The final isothermal phase of the equation of state is simply in order to allow
sink-particle formation to occur,  which requires a subvirial collapsing fragment.

Star formation in the cloud is modelled through the introduction of sink-particles (Bate, Bonnell \& Price 1995).
Sink-particles formation is allowed once  the gas density of a collapsing fragment reaches $\rho\ge 6.8 \times 10^{-14}$ g cm$^{-3}$ although the equation of state ensures that this requires $\rho\ge 2. \times 10^{-13} {\rm g cm}^{-3}$. The neighbouring SPH
particles need be within a radius of $1. \times 10^{-3}$ pc and that fragment must be subvirial and collapsing.
Once created, the sinks accrete bound gas within $1. \times 10^{-3}$ pc and all gas that comes within $2. \times 10^{-4}$ pc.
The sinks have their mutual gravitational interactions smoothed to $2. \times 10^{-4}$ pc or 40 au. No interactions
including binary or disc disruptions can occur within this radius. 

\section{brown dwarf formation in turbulent molecular clouds}

The simulation was followed for $1.02$ free-fall times or $\approx 6.6 \times 10^5$ years
and $\approx 3.9\times 10^5$ years after the first stars formed. During this time, 2542 stars
were formed with masses between $0.017$ and 30 \solmas. Of these, $\approx 23$ per cent
have masses below $0.08 \solm$ but only $ \approx 10$ per cent  (243 objects)
have $m\le 0.08 \solm$ and have stopped accreting. A further 3 per cent  are likely
to maintain $m\le 0.08 \solm$ given their final accretion rates and assuming that
this accretion is sustained over the next free-fall time. This gives an expected final
number of brown dwarfs of 342  or 13 per cent of the stars formed.
Figure~\ref{BDsimage} shows the spatial distribution of the brown dwarfs at the
end of the simulation. They are primarily in or around forming stellar clusters. We will investigate the
process by which they form in the following sections.

\begin{figure*}
\centerline{\psfig{figure=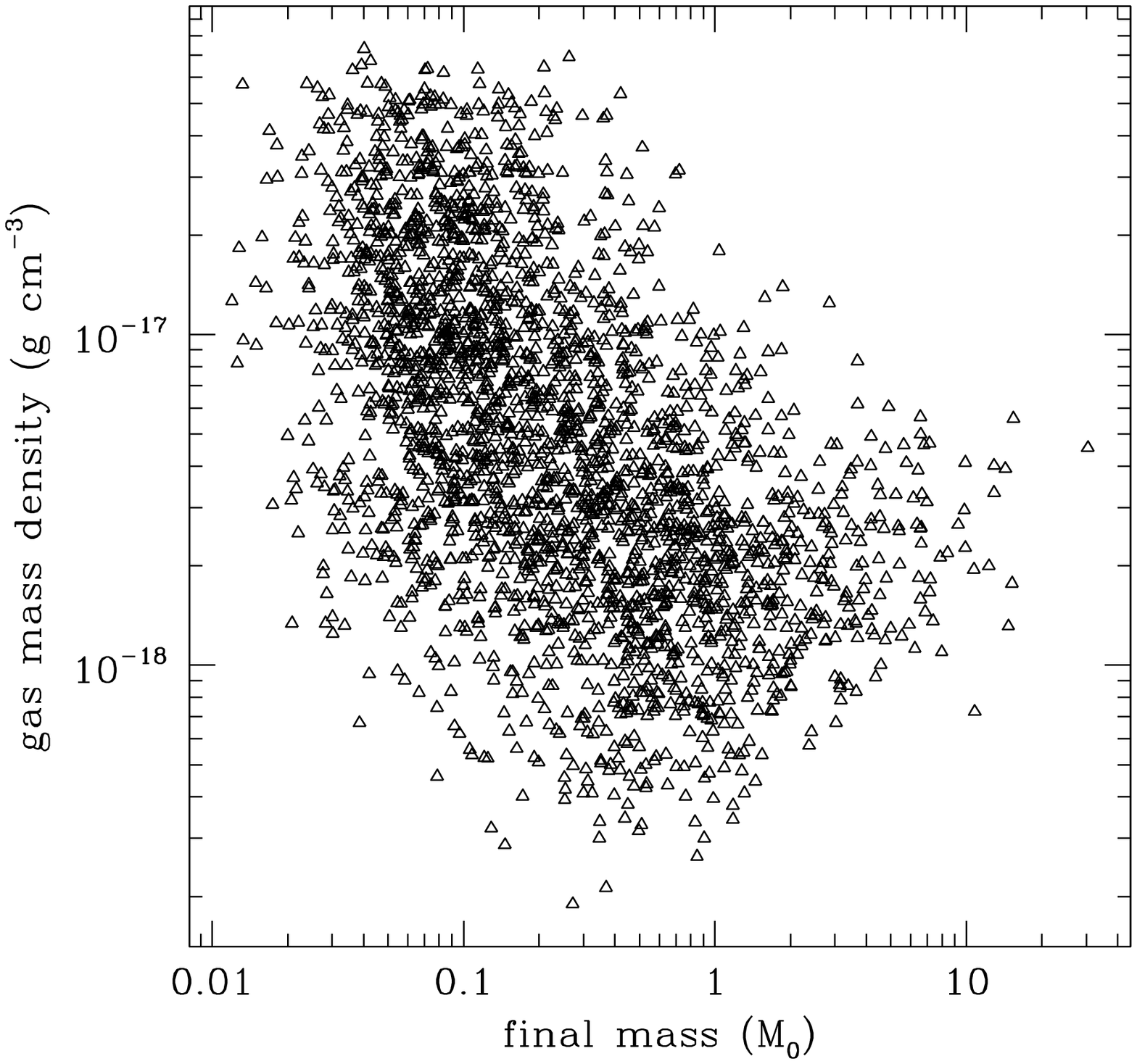,width=8.truecm,height=8.truecm,rheight=8.truecm}
\psfig{figure=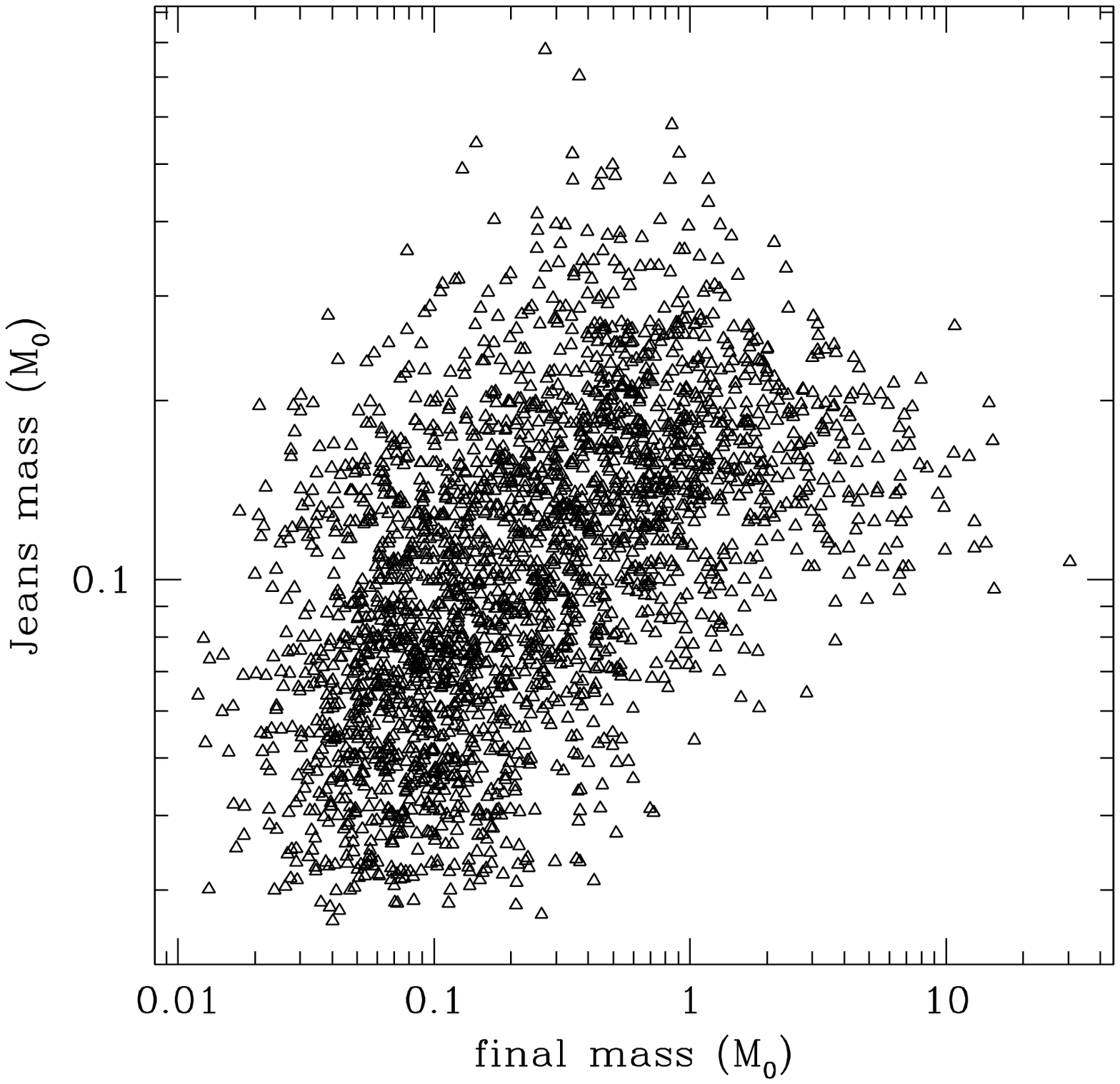,width=8.truecm,height=8.truecm,rheight=8.truecm}}
\caption{\label{bdformgasden} The gas density (left) and Jeans mass (right) are plotted against the final stellar masses
for the 2542 sinks formed. Both the gas density and Jeans mass are calculated within a radius of $0.05$ pc from the
incipient sink particle within $\approx 2300$ years prior to the sink formation. The gas densities are higher, and the Jeans masses
lower, for the low-mass sinks formed as is expected when the physical conditions of the gas determine the fragment masses. }
\end{figure*}

Most formation mechanisms for brown dwarfs envision that the physical conditions
in the pre-fragmented gas are such that the Jeans mass, the minimum 
mass to be gravitationally bound, be of order a brown dwarf mass (Elmegreen 2004; Whitworth \etal\  2007; Bonnell \etal\ 2007). 
Figure~\ref{bdformgasden} plots the median gas density, and 
respective Jeans mass, within  $0.05$ pc (a typical Jeans length) of where the sink-particle will form. These values
are calculated from the gas distribution just prior (within 2300 years) to the formation of the sink-particle.
The gas densities and Jeans masses are plotted against the final masses that these sinks
attain by the end of the simulation. We see that the moderate mass sinks ($0.5-2 \solm$) form in low density gas
where the Jeans mass is of order $0.1$ to almost $1.0$ \solmas. In contrast, the lower-mass sinks form
from higher density gas and thus from low Jeans masses. Note that there is not a perfect one-to-one correspondence
between the Jeans masses and the final masses even for small masses due to the somewhat arbitrariness
of evaluating the physical conditions within a fixed radius of $0.05$ pc. What is important is that
the physical properties in the gas are appropriate to give very low-mass fragments such that
forming brown dwarfs in these regions is natural.
Subsequent accretion can
also increase the final masses from the low value generated at the point of the fragmentation. This
is more evident for the high-mass sinks where the vast majority of their final masses comes
from the subsequent competitive accretion (Bonnell \etal\ 2001; Bonnell, Vine \& Bate 2004; Bonnell \& Bate 2006).

The first conclusion we can make is that the brown dwarfs form as lower-mass stars do, from the fragmentation
of a gas cloud where the thermal Jeans mass is of order the mass of the object formed. The next question is what
drives the gas to such densities that brown dwarfs can form. One possibility is that turbulent compression
leads to the formation of dense cores and thus brown dwarfs (Padoan \& Nordlund 2004). The spatial
distribution of the brown dwarfs in figure~\ref{BDsimage} argues against this as the brown dwarfs are not
located randomly throughout the cloud but are located in the vicinity of stellar clusters. Furthermore,
as SPH is Lagrangian, we can trace the  particles that form individual sinks backwards in time.
We can estimate the fraction of an individual sink's
mass that is brought to the point of formation by the turbulent flows. This shows that 
the turbulent flows are responsible for transporting only of order one per cent of the fragment's
mass to within 0.04 pc  of  where the sink eventually forms. Instead additional acceleration,  such as occurs when the gas enters the gravitational potential of the cluster, are necessary.  We can therefore conclude that, in {\sl this simulation}, turbulence does not
lead to the formation of brown dwarfs.

\begin{figure}
\centerline{\psfig{figure=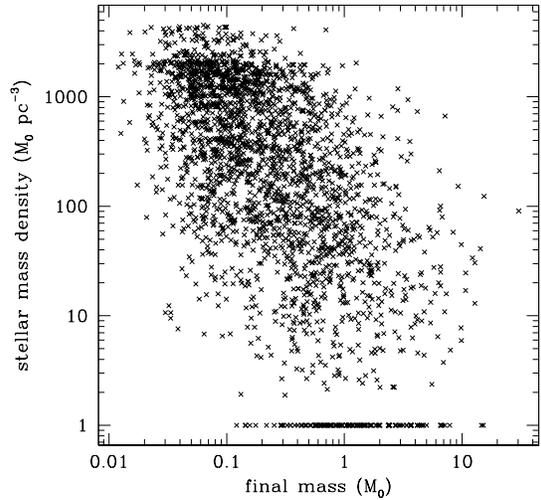,width=8.truecm,height=8.truecm,rheight=8.truecm}}
\caption{\label{stellclusts} The mass density of stars within $0.25$ pc of the forming sink-particle is plotted as a function of the final mass in the sink-particle. The moderate and high-mass sinks form in regions of low stellar mass density whereas
the low-mass stars and brown dwarfs appear to require the presence of a stellar cluster and its strong gravitational potential in order
to form. The gravitational potential of the cluster generates the high gas densities needed for forming low-mass objects.  The stellar mass densities are calculated within $\approx 2000$ years prior to the sink formation.  Regions completely devoid of
stars prior to the sink formation are given a stellar mass density of $1$ \solmas\ pc$^{-3}$. Note that although the high-mass stars form in regions of low  stellar mass density, they end up in the centre
of the clusters which form subsequently around them.}
\end{figure}

\section{Brown dwarf formation in stellar clusters}

Gravity provides an alternative to turbulence in generating  the high gas density conditions conducive to forming brown dwarfs. Gravity has the distinct advantage over turbulence in that it is intrinsically convergent and
therefore better able to compress gas into small volumes. It also has the tendency of compressing a three-dimensional
volume into two-dimensional sheets, one dimensional filaments and eventually to point sources (Larson 1985; Bonnell 1999).
The stellar clusters forming in the simulation provide a strong gravitation potential into which flowing gas
can be compressed to much higher densities.  Figure~\ref{stellclusts} shows the stellar mass density
within $0.25$ pc, a typical size-scale of a stellar cluster,  of the forming sink-particles just prior to sink formation. We see that while the moderate
and high-mass sinks form in regions where the mass density of stars is low, low-mass stars and brown dwarfs form
in regions of high stellar mass density. This occurs as the moderate to higher-mass stars form first while the
eventual low-mass stars and brown dwarfs only form as gas infalls into pre-existing stellar clusters. 
It is therefore the presence of the stellar cluster which is driving the formation of these objects. 
The higher-mass stars form the basis for the forming cluster and thus also end up inside a region of high stellar mass density.


\begin{figure}
\centerline{\psfig{figure=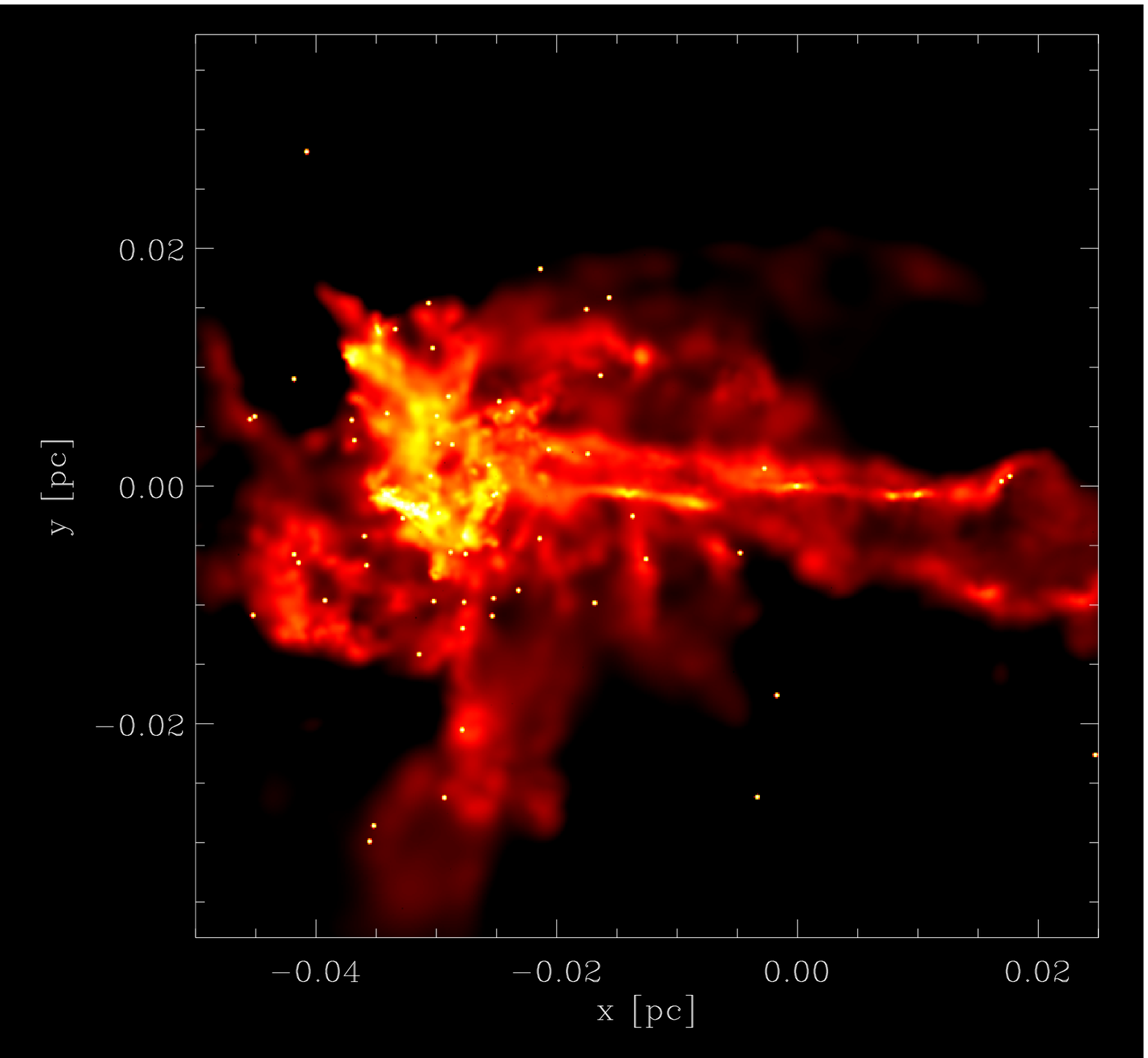,width=8.truecm,height=8.truecm,rheight=8.truecm}}
\centerline{\psfig{figure=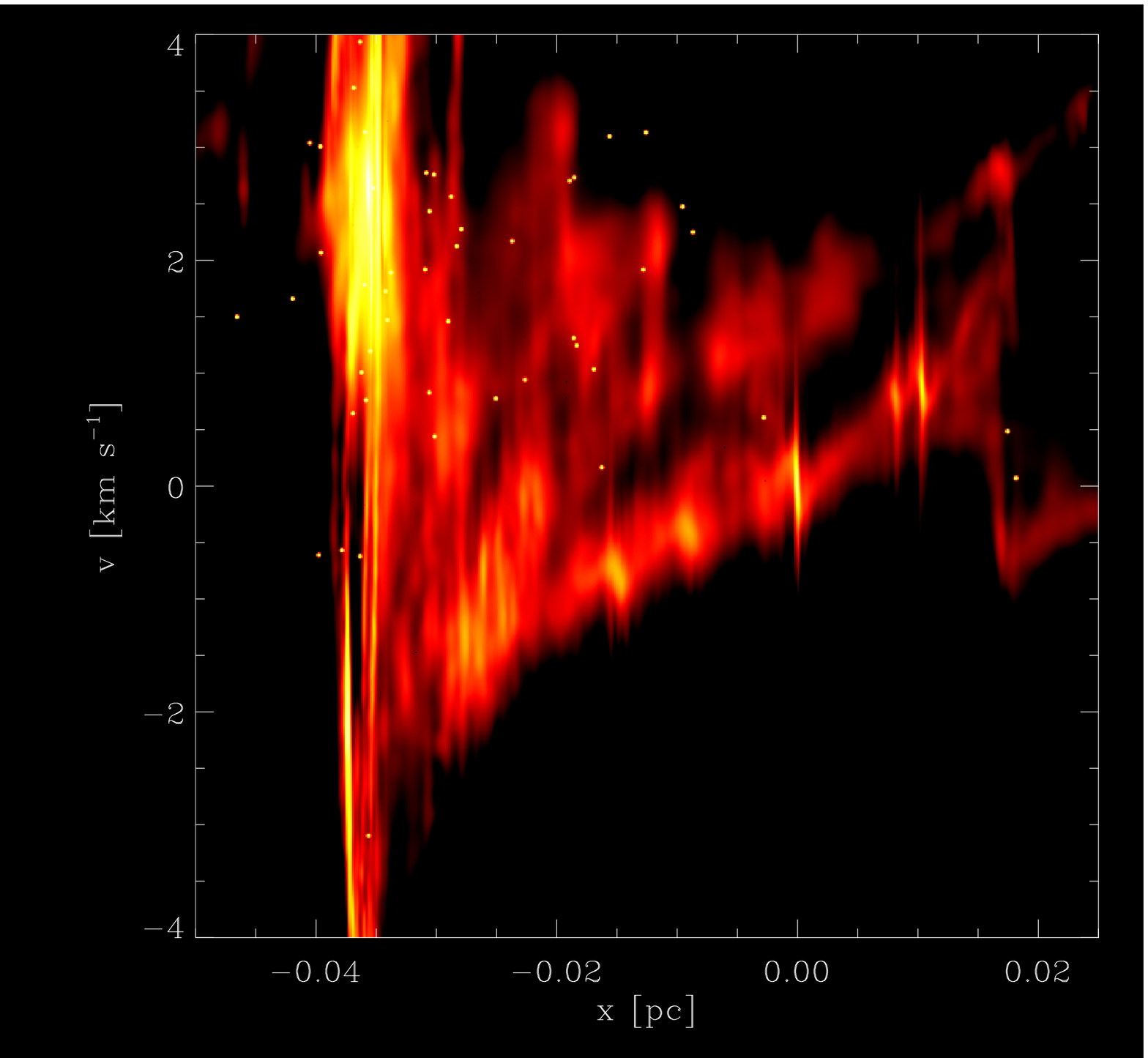,width=8.truecm,height=8.truecm,rheight=8.truecm}}
\caption{\label{bdformfil} 
In the top panel we show a column density image of a gaseous filament which is forming brown dwarfs as it accretes onto a rich cluster. The position (0,0) in the co-ordinates shows the formation site of one of the new brown dwarfs. The column density scale runs from 0.5 to 50 g cm$^{-2}$. The bottom panel shows a position-velocity diagram for the same region where the x-positions are centred on the same brown dwarf as in the upper panel. The velocities are calculated projected along the vector joining the brown dwarf at position $x=0$ to the centre of mass of the cluster, and are again centred on the brown dwarf. Note that the colours in the bottom panel  scale from 5 to 500 g cm$^{-1}$ km$^{-1}$ s. The diagonal form of the filament in position-velocity diagram reveals that the gas is accelerating into the cluster while being tidally sheared away from the objects that are forming within. Relative to the formation site of the brown dwarf, the vast majority of the filament is moving away: gas at negative positions have negative relative velocities, while gas a positive positions have positive relative velocities. Only in the immediate region surrounding the still-forming brown dwarf can one see a reversal of this velocity signature, which denotes the gas falling on to the new object. This feature can be seen in several other points along the filament, showing the formation sites of other brown dwarfs or low-mass stars.}
\end{figure}

Figure~\ref{bdformfil} shows the spatial distribution of the
gas and stars within $0.05$ pc of a forming brown dwarf. We see a large filament formed through
the amplification of structure in the infalling gas due to the combination of the cluster potential and its self-gravity. 
Several knots of gas are forming along the filament, some of which go on to form brown dwarfs while
the others collapse to form low-mass stars. Such an occurrence of a filamentary structure infalling into the
cluster is relatively common, although often some degree of tangential motion and thus angular momentum
result in the filament being wound up around the protocluster as it infalls and forms low-mass objects.

\begin{figure}
\centerline{\psfig{figure=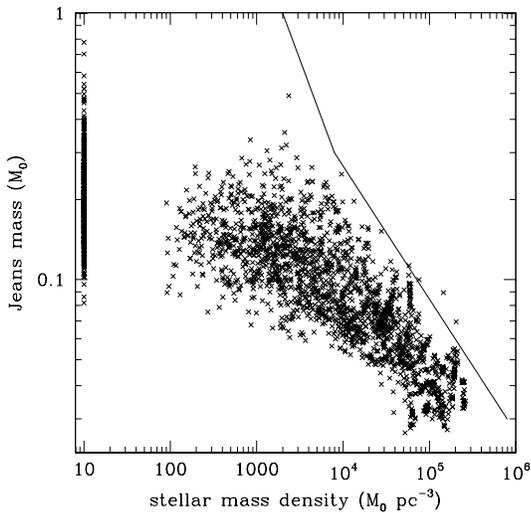,width=8.truecm,height=8.truecm,rheight=8.truecm}}
\caption{\label{ClustJeans} The Jeans mass within $0.05$ pc of a proto-sink is plotted against the stellar mass density
in the same region for the 2542 sinks formed. Note that the Jeans mass declines with increasing stellar mass density.
This occurs as the gas density is higher in stellar clusters due to the effects of the gravitational potential
and also that lower gas density fragments would be tidally disrupted and hence not form any sinks. The solid
line indicates the Jeans mass assuming a gas density that is the same as the stellar mass density. This
tidal limit provides an upper limit on the possible Jeans masses in the stellar cluster. Both the stellar
mass densities and Jeans masses are calcualted  within $\approx 2300$ years prior to the sink formation.  The Jeans
mass is based on the median gas density in the region.}
\end{figure}

Figure~\ref{bdformfil} also plots the associated position-velocity diagram of the infalling filament  centred on the position and velocity of
one of the forming brown dwarfs.  The negative slope shows that the filament is falling into the 
cluster located at $\approx -0.035$ pc and $\approx + 2$ km s$^{-1}$. The gas is accelerated
into the cluster such that material at negative (positive)  positions relative to the forming brown dwarf
have negative (positive) relative velocities, respectively. This divergent flow, as well as the large
virial velocity the forming brown dwarf receives due to the cluster potential, restricts any subsequent accretion and thus maintains the low-mass of the object. 
On smaller scales, gravitational collapse reverses this pattern
in the position velocity-diagram. Material at negative positions has positive relative velocities while material at positive
positions has negative relative velocities indicating collapse.

This tidal shearing of the filament acts to limit the  mass of the 
object. Gas that would normally fall into the forming fragment is now pulled away by the large-scale potential. 
As a result, only the  very high gas density
fragments, and hence with low Jeans masses, form in such a shear flow. Figure~\ref{ClustJeans} plots the Jeans mass
of the 2542 sinks at the point of formation as a function of the stellar mass density within $0.05$ pc. We see that the
Jeans masses decrease as the cluster stellar mass density increases indicating the prevalence for forming low-mass 
objects in dense stellar clusters.  Also plotted in figure~\ref{ClustJeans} is the Jeans mass if the gas density was equal to the
stellar mass density. This is  the tidal limit for forming a fragment and provides an upper limit to the distribution
of Jeans masses of the forming fragments. Higher-mass fragments cannot form as their high Jeans masses and low gas densities
would result in their being tidally disrupted before they could collapse.

The growth of a stellar cluster, acts to continually decrease the Jeans mass and thus
the fragmentation mass in the infalling gas. As more stars form or fall into the cluster, the increase
in the stellar density increases the tidal shearing and thus necessitates higher gas densities
in order for the fragments to be bound. Thus the fragment mass should decrease with time and with the growth of a cluster. Subsequent accretion onto some of these objects will significantly increase their masses.

\section{cluster dynamics and accretion}

\begin{figure}
\centerline{\psfig{figure=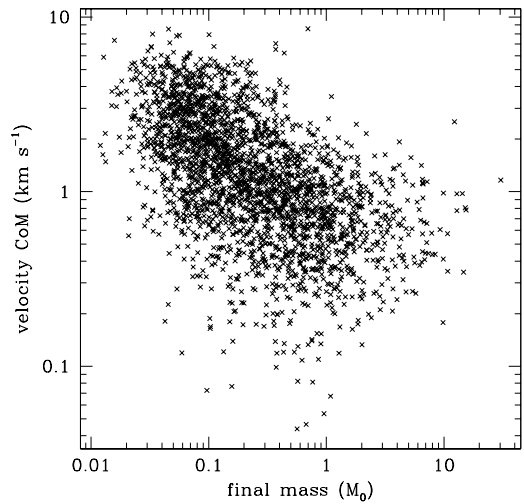,width=8.truecm,height=8.truecm,rheight=8.truecm}}
\caption{\label{velcom} The relative velocity of the 2452 sinks at formation relative to the centre
of mass velocity within $0.25$ pc is plotted against the final mass of each sink. The sinks
that  attain moderate or high masses form with low relative velocities to their environment while those
that remain low-mass objects form with high velocities relative to the centre of mass of their environments.
This occurs as the sinks that attain moderate or high masses form earlier and constitute the initial stars in the
cluster. The low-mass sinks form from gas falling into a already formed stellar cluster and thus have
a high relative velocity. This high velocity also limits any subsequent accretion ensuring they remain low-mass
objects.}
\end{figure}

In the previous section we saw that the low-mass sinks form as the gas infalls into an already formed stellar cluster.
This has important implications as to why they remain low-mass objects of the typical mass of a brown
dwarf rather than accreting from the abundant reservoir of gas in the stellar cluster. Figure~\ref{bdformfil}
shows the gas being accelerated to high (negative) velocities as it infalls into the cluster's gravitational potential. 
Thus any newly formed fragment will enter the cluster at high velocity and therefore have difficulty in accreting from
the reservoir of gas in the cluster. Figure~\ref{velcom} shows the relative velocity of each forming sink compared to its
environment within $0.25$ pc.  The sinks that remain low-mass have high relative velocity, of order several km s$^{-1}$ at the point of formation. In contrast, 
the sinks that ultimately attain moderate and high mass form with low relative velocities. This difference occurs
as the low-mass sinks form from gas infalling into an already formed cluster whereas the sinks that attain high
mass form before any stellar cluster is present. They are the first stars around which the cluster is built
and due to their low relative velocity, they are able to accrete significant amounts of gas and become high-mass objects. 
The low-mass sinks cannot accrete significantly due to their high relative velocity and thus remain low-mass objects.
Estimated accretion rates for a low mass star ($m\le 0.1 \solm$) travelling at a high velocity ($v\ge 2$ km s$^{-1}$) 
in a gas reservoir of $10^{-17}$ g cm$^{-3}$ is $\approx 5 \times 10^{-8}$ \solmas\ yr$^{-1}$,  which is too
small to significantly alter the star's final mass over accretion times of $t_{\rm acc} \simless 10^6$ years (c.f. Bonnell \& Bate 2006).

In this scenario, the forming brown dwarfs and low-mass stars do not require any subsequent interactions or ejections to terminate their
accretion processes. It is the acceleration due to the cluster potential which ensures that they have low accretion rates
and remain low-mass objects. Brown dwarfs and low-mass stars thus form in the same way from low Jeans mass fragments which
subsequently accrete little mass. If ejections are not required, then the disc and binary properties of these objects need
not be affected and should form a continuous distribution with slightly higher-mass stars.

\section{Observational signatures}

\begin{figure}
\centerline{\psfig{figure=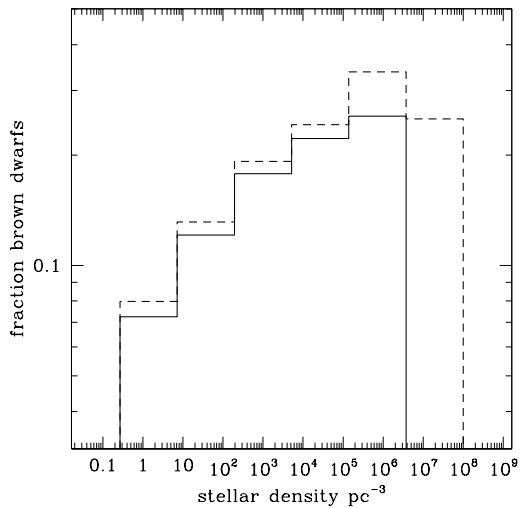,width=8.truecm,height=8.truecm,rheight=8.truecm}}
\caption{\label{fracbds} The fractional abundance of brown dwarfs is plotted against their local
stellar density based on their tenth closest neighbour. The dashed line is for all brown-dwarf-mass sink-particles
at the end of the simulation while the solid line only considers brown dwarf-mass sink-particles
with low accretion rates such that they would remain brown dwarfs if they continued to accrete over the
following free-fall time ($6.5 \times 10^5$ years).  The fractional abundance
of brown dwarfs is low at low stellar densities and peaks at high stellar densities inside 
clusters. The highest stellar densities also appear to have lower fractional abundances of brown dwarfs.}
\end{figure}

The brown dwarfs that form in the scenario described here form in an analogous manner as do very low-mass stars (Elmegreen 2004).
There is nothing specific to being a brown dwarf that makes their formation distinct from stars. They form
from gas which has low Jeans masses due to its compression as it enters a stellar cluster, and subsequent accretion
is generally low due to the high infall velocity imparted from the cluster potential. There is no need for
close interactions or ejections to ensure their low mass. This implies that the properties of brown dwarfs will generally
resemble those of low-mass stars. As such, their circumstellar disc and multiplicity properties should form a 
continuum with low-mass stars. This process can occur in any strong potential such 
as that of a small-N cluster where 
the Jeans mass becomes small and that the virial velocity limits subsequent accretion
(Bonnell \etal 1997,~2001; Bate \etal 2002; Bonnell \& Bate 2006). Ejections can occur but are
not fundamental to the process.

The velocities of the brown dwarfs and other low-mass objects are also similar to those of the
other stars as both simply reflect  
the overall gravitational potential. Although the brown dwarfs form at higher relative velocities
than do the higher mass objects, this is simply due to the presence
or absence of a well defined gravitational potential at the point of formation. The velocities of the higher mass objects are initially low as they form before the cluster develops. Their velocities      
increase over time due to the infall of other stars into the growing cluster. 
Virial equilibrium ensures that they all  have similar velocity dispersions.

One significant observable feature is that the brown dwarfs, and for that matter the very low-mass stars, require a gravitational
potential to compress the gas to sufficiently high densities to attain low Jeans masses. This means that they form in regions of high
stellar densities as shown in figure~\ref{stellclusts}. There is some subsequent dynamical evolution and brown dwarfs, like low-mass stars,
are more likely to be ejected in any dynamical event. Nevertheless, the brown dwarfs formed are more commonly found in a clustered
environment. Figure~\ref{fracbds} shows the fractional abundance of brown dwarfs as a function of stellar density. The stellar density is calculated from the distance to the ten nearest neighbours of each sink-particle. The abundance of brown dwarfs is plotted for all sinks
that have masses in the range a brown dwarf masses and also for those that given their final accretion rates are likely to remain brown dwarfs over the following free-fall time, $\approx 6.5 \times 10^5$ years. We see that the frequency of brown dwarfs is significantly higher
in the clustered than in the non-clustered regions, with a peak abundance of $\approx 25$ per cent at high stellar densities 
which decreases to $\simless 10$ per cent in isolated or regions of low stellar densities. 

In general, the brown dwarf mass sinks in the lower density bins are those that have been ejected from their natal clusters.
This small number is thus likely to evolve with time from basically zero at the point of formation, and should increase somewhat
over our final value as more low-mass cluster members are ejected from the clusters. Nevertheless, we expect that an observable
signature of the process described here is that the fractional abundance of brown dwarfs should increase with stellar density, and should
be lowest amongst the isolated population of young stars. It is worth noting that the fractional abundance of brown dwarfs actually decreases in the highest stellar density bins as the stars in the cores of the stellar clusters are predominantly higher mass stars due
to the ongoing (competitive) accretion there.

\section{Conclusions}

We have investigated the formation of brown dwarfs in a numerical simulation of a self-gravitating turbulent molecular cloud. We find that the brown dwarfs form, as do low-mass stars, due to the
fragmentation of high-density gas that arises as it infalls into the gravitational potential
of a stellar cluster. 
Approximately 23 per cent of the objects formed have final masses in the brown dwarf mass range although only $\approx 10$ per cent have brown dwarf masses and have stopped mass accretion. 

The turbulent velocities present in the cloud do not contribute directly to the formation of the
brown-dwarf mass fragments. Instead, these fragments form  through the gravitational compression of gas as it infalls into a stellar
cluster. This intrinsically 3-D compression produces high gas densities and thus low thermal Jeans masses in the range of brown dwarfs and low-mass stars. 
The tidal shear of the cluster, and the velocity imparted
on the fragment from the cluster potential act to limit any subsequent mass increase due to accretion.
There is no need for any subsequent ejections to halt the accretion implying that the
circumstellar disc and binary properties of brown dwarfs should form a continuum with 
low-mass stars. 

Brown dwarfs formed through this mechanism should be preferentially located in regions of high stellar density. The fractional abundance of brown dwarfs in stellar clusters is of order 25 per cent in highly
clustered regions whereas it decrease to of order 10 per cent in isolated regions. This fraction
is likely to increase somewhat due to subsequent ejections of brown dwarfs and low mass stars from
the clusters.

\section*{Acknowledgments}
We acknowledge the contribution of the  U.K. Astrophysical Fluids Facility (UKAFF) and SUPA
for providing the computational facilities for the simulations reported here. P.C.C. acknowledges support by the Deutsche Forschungsgemeinschaft (DFG) under grant KL 1358/5 and via the Sonderforschungsbereich (SFB) SFB 439, Galaxien im fr\"uhen Universum. 
MRB is grateful for the support of a Philip Leverhulme Prize and a EURYI Award.
This work, conducted as part of the award ÔThe formation of stars
and planets: Radiation hydrodynamical and magnetohydrodynami-
cal simulationsÕ made under the European Heads of Research Coun-
cils and European Science Foundation EURYI (European Young
Investigator) Awards scheme, was supported by funds from the Par-
ticipating Organizations of EURYI and the EC Sixth Framework
Programme.
We would like to thank Chris Rudge and Richard West at the UK Astrophysical Fluid Facility (UKAFF) for their tireless assistance and enthusiasm during the completion of this work. Finally, we thank  the referee
for some useful comments which helped clarify the text.

\end{document}